\documentclass{resonance}

\usepackage{amsfonts}
\usepackage{amsmath}
\usepackage{amssymb}
\usepackage{color}
\def\i{\item}

\newcommand{\bed}{\begin{displaymath}}
\newcommand{\eed}{\end{displaymath}}
\newcommand{\bei}{\begin{itemize}}
\newcommand{\eei}{\end{itemize}}
\newcommand{\bef}{\begin{figure}}
\newcommand{\eef}{\end{figure}}
\newcommand{\ben}{\begin{enumerate}}
\newcommand{\een}{\end{enumerate}}
\newcommand{\beq}{\begin{equation}}
\newcommand{\eeq}{\end{equation}}
\newcommand{\ber}{\begin{eqnarray}}
\newcommand{\eer}{\end{eqnarray}}

\newcommand{\bb}{\bf B}

\newcommand{\jb}{\bf J}

\newcommand{\gcc}{\mbox{${\rm g} \, {\rm cm}^{-3}$}}

\newcommand{\msun}{\mbox{{\rm M}$_{\odot}$}}

\newcommand{\lsim}{\raisebox{-0.3ex}{\mbox{$\stackrel{<}{_\sim} \,$}}}

\newcounter{attnctr} 

\definecolor{cblue}{rgb}{0.9,0.9,1.0}
\definecolor{darkblue}{rgb}{0.1,0.1,0.6}
\definecolor{darkred}{rgb}{0.6,0.1,0.1}
\begin{document}


\title{Gravity Defied \\
     From potato asteroids to magnetised neutron stars}
\secondTitle{III. White Dwarfs (dead stars of the first kind)}
\author{Sushan Konar}

\maketitle
\authorIntro{\includegraphics[width=2cm,angle=90]{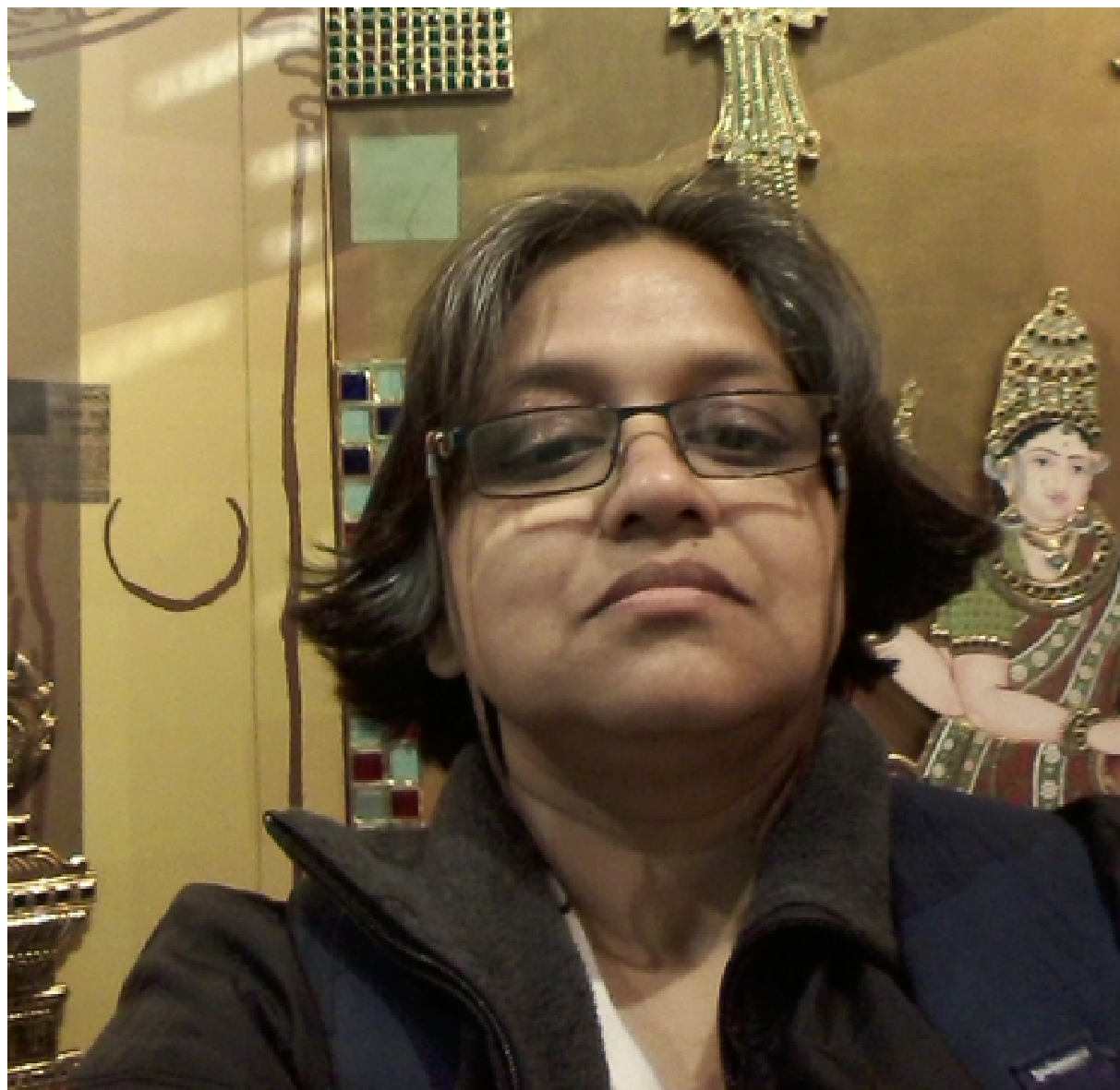}\\
Sushan Konar  works at  NCRA-TIFR, Pune. She  tries to  understand the
physics of stellar  compact objects (white dwarfs,  neutron stars) for
her livelihood and writes a blog about life in academia as a hobby.}
\begin{abstract}
  During  its  active lifetime  a  star  burns  its nuclear  fuel  and
  gravitation is held off by the  pressure of the heated gas.  Gravity
  should  take over  once this  fuel  is exhausted  unless some  other
  agency saves the star from such a fate. Low mass stars find peace as
  {\bf  \em white  dwarfs}  when  the electrons  settle  into a  Fermi
  degenerate phase where the  pressure of degenerate electrons balance
  the gravitational pressure.
\end{abstract}
\monthyear{2017}
\artNature{GENERAL  ARTICLE}

\section{The Stars}

The nature  of the stars  has been  questioned and debated  over ever
since  the dawn  of human  intelligence. Yet,  it's only  in the  late
nineteenth century when  the Sun and other stars  have been understood
to be  self-gravitating gaseous objects. We  now know that the  Sun is
powered by nuclear fusion, producing  Helium from Hydrogen, the direct
evidence of  which has come from  the detection of solar  neutrinos in
1968.  In  fact, it is  this particular nuclear reaction  that defines
the birth  of a star.  When  a self-gravitating gas cloud  attains the
capability of fusing Hydrogen into Helium, a star is born!

\begin{figure}[!t]
  \caption{Pillars of molecular clouds in  Eagle nebula - a birthplace
    of stars.  Image courtesy - {\tt https://jwst.nasa.gov/}}
  \label{f_eagle}
\vspace{-0.5cm} 
\centering\includegraphics[width=5.0cm]{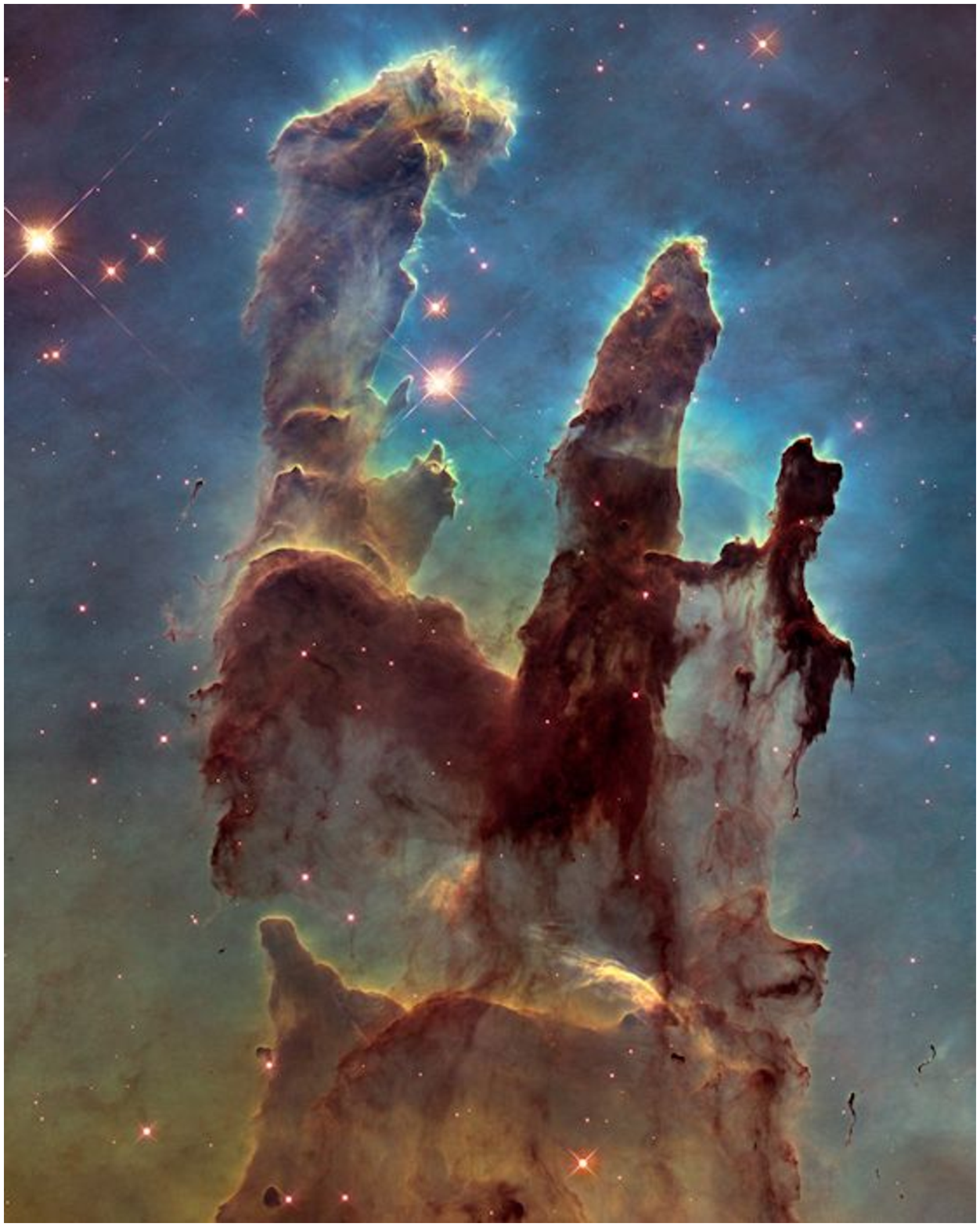}
\vspace{-0.5cm} 
\end{figure}

Giant molecular clouds in the  interstellar medium are the birthplaces
of new  stars (Fig.~\ref{f_eagle}).  The interstellar  medium consists
  mainly  of atomic  hydrogen ($\rho  \sim 0.1$~atoms/cm$^3$,  $T \sim
  10^4$~K).  This exceptionally diffuse gas hosts two types of massive
  clouds ($10^7$~\msun)  of denser  gas of which  the more  dense ones
  consist  almost entirely  of  molecular  hydrogen.  Supernova  blast
  waves,  impacting upon  these molecular  clouds, create  shock waves
  heating the clouds  up to $T \sim 10^6$~K  (temperature required for
  Hydrogen fusion) which ultimately trigger star formation.

However,  before actual  star formation  Jeans instability  causes the
collapse and  subsequent fragmentation of these  molecular clouds into
star-sized  clumps.   We  have  seen earlier  how  the  condition  for
hydrostatic equilibrium depends on  the temperature ($T$) (responsible
for the gas  pressure) and the mass  ($M$) of the gas.   The gas cloud
collapses   if  its   temperature   is  not   sufficient  to   balance
gravitational pressure.  Conversely, for  a given temperature, the gas
would  be stable  for  sufficiently  small masses  but  would begin  a
process of runaway contraction when a critical mass is exceeded.

This critical  mass, known as the  Jeans mass (after Sir  James Jeans)
can be  estimated as  follows. Consider an  isothermal gravitational
contraction of  a homogeneous spherical  gas cloud of mass  $M$, radius
$r$  which is  at a  temperature $T$.   When the  gravitational energy
release due to contraction exceeds the  work done on the gas cloud, we
have an episode of runaway  contraction.  Therefore, the critical mass
is  obtained  when  the  work done  equals  the  gravitational  energy
release.  The gravitational energy release  from the gas cloud when it
contracts from a radius $r$ to $r-dr$ is
\beq
  dE_{\rm G} = G \frac {M^2}{r^2} dr\,,
\eeq
and the work done on the gas cloud during this process is
\beq
  dW \propto \rho T r^2 dr \,,
\eeq
where $\rho$  is the density  of the gas and  the cloud is  assumed to
behave like an idea gas ($ P  \sim \rho T$). Equating $dE_{\rm G}$ and
$dW$ we arrive at the Jeans mass given by
\beq
M_{J} \propto \left({\frac {T^{3}}{n}}\right)^{\frac {1}{2}} \,.
\eeq
It  should  be  noted  here  that the  above  formulation  assumes  an
isothermal contraction.  For adiabatic  processes the Jeans mass turns
out to be
\beq
M_{J}^{a} \propto \rho^{\frac{3}{2}(\gamma - \frac{4}{3})}\,,
\eeq
where  $\gamma$  is  the  adiabatic  index of  the  gas.   It  can  be
immediately  seen that  $M_J$  increases with  increasing density  for
$\gamma >  4/3$, and decreases  with increasing density for  $\gamma <
4/3$.

\begin{figure}
  \caption{Binding energy ($E_B$) per nucleon in stable atomic nuclei.
    Fe$^{56}$ has the minimum  binding energy per nucleon.  Therefore,
    beyond  Fe$^{56}$, fusion  of elements  no longer  release energy.
    Stellar fuel is exhausted once Fe$^{56}$ is made.}
  \label{f_binding}
\vspace{-0.5cm} 
\centering\includegraphics[width=8.5cm]{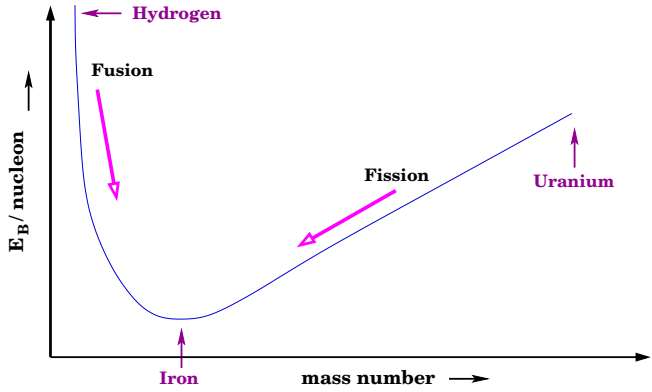}
\vspace{-0.5cm} 
\end{figure}

In summary  then the  path taken  by a giant  molecular cloud  to star
formation is as follows -
\bei
   \vspace{-0.5cm}
   \i Giant  molecular clouds  of gas  and dust begin  with a  mass of
   $\sim  10^3 -  10^6$~\msun~and an  initial composition  of Hydrogen
   with  a  small  mixture   of  Helium,  molecular  Hydrogen,  water,
   silicates..
   \vspace{-0.5cm}
   \i The  clouds collapse  gravitationally and fragment  into clumps.
   The small clouds stick together and grow through accretion, and are
   compressed by supernova blast waves.
%
   \i The clumps collapse and fragment further to form proto-stars.
%
   \i Protostars collapses till hydrogen fusion starts at the core.
\eei

For most of a star's life the gravitational force and the gas pressure
balance each other. This  balance is finely-tuned and self-regulating:
if the rate of energy generation  in the core slows down, gravity wins
out over pressure  and the star begins to  contract.  This contraction
increases the temperature and pressure  of the stellar interior, which
leads to higher energy generation rates and a return to equilibrium.

\begin{figure}
  \caption{The evolutionary path and the final remnant (white dwarf
    / neutron star / black hole) of a star depends crucially on its
    initial mass. Legends : MS - main sequence, RG - red giant, SN - supernova}
  \label{f_rmnt}
\vspace{-0.5cm} 
\centering\includegraphics[width=9.0cm]{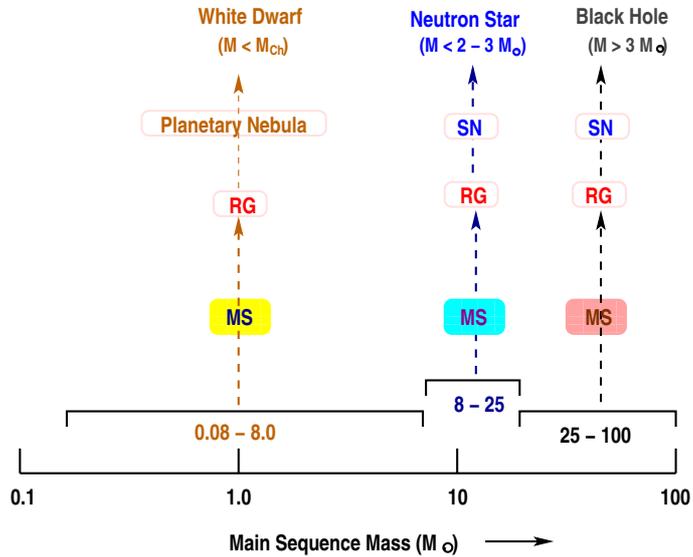}
\vspace{-0.5cm} 
\end{figure}

Once formed,  the stars spend most  of their life fusing  Hydrogen, in
the  phase known  as {\em  main sequence}.  During main  sequence, the
stars    get    progressively     hotter    and    brighter    (higher
luminosity).  Depending  on  the  mass   the  stars  then  enter  into
subsequent  phases of  nuclear  fusion involving  Helium and  elements
upwards of  Helium.  Fusion  is an  exothermic process  that generates
energy - when two nuclei of  lighter mass fuse into a heavier nucleus,
the mass per nucleon actually decreases  and this mass defect shows up
as  energy.  The  mass per  nucleon keeps  decreasing with  increasing
atomic  number  till  Fe$^{56}$ (Fig.~\ref{f_binding}).   Beyond  this
point, mass per  nucleon increases and fusion can only  be achieved by
supplying energy from outside.  Therefore,  the nuclear source of fuel
is completely exhausted upon production of Fe$^{56}$.

Therefore, the ultimate phase of  stable equilibrium of a star depends
on its initial  mass. In this phase gravitation is  stably balanced by
electromagnetic  or   quantum  effects  and  there   is  no  evolution
hereafter.  The   evolution  and  the  end-states   are  described  in
Fig.~\ref{f_rmnt} and Table~\ref{t_rmnt}.

\begin{table}[!h]
  \caption{The end state of a star as a function of its mass at the
    beginning of the main-sequence ($H \rightarrow He^4$) phase.
    Legend : WD - white dwarf}
\label{t_rmnt}
\vspace{-0.5cm}
\centering
{\small
\begin{tabular}{lll}
{\bf Main-Sequence Mass}       & {\bf Final State}       & {\bf \em gravity-resisting agent}\\
$     \, \lsim 0.01 \, \msun           $  & Planet       & van der Waals \\
$0.01 \, \lsim M/\msun \, \lsim 0.08   $  & Brown Dwarf  & Fermi degeneracy \\
$0.08 \, \lsim M/\msun \, \lsim 0.5    $  & He WD        & - do - \\
$0.5  \, \lsim M/\msun \, \lsim 8      $  & C-O WD       & - do - \\
$8    \, \lsim M/\msun \, \lsim 10     $  & O-Ne-Mg WD   & - do - \\
$10   \, \lsim M/\msun \, \lsim 25 - 40$  & Neutron Star & - do - \\
$40   \, \lsim M/\msun \,              $  & Black Hole   & {\bf \em none} \\
\end{tabular}
}
\end{table}

\section{White Dwarfs}

\begin{figure}[!b]
  \caption{NGC 2440, a planetary nebula, contains one of the hottest white
    dwarfs (seen as a bright dot near the centre). Image courtesy -
    {\tt https://www.nasa.gov/}}
  \label{f_pnebula}
\vspace{-0.5cm} 
\centering\includegraphics[width=7.5cm]{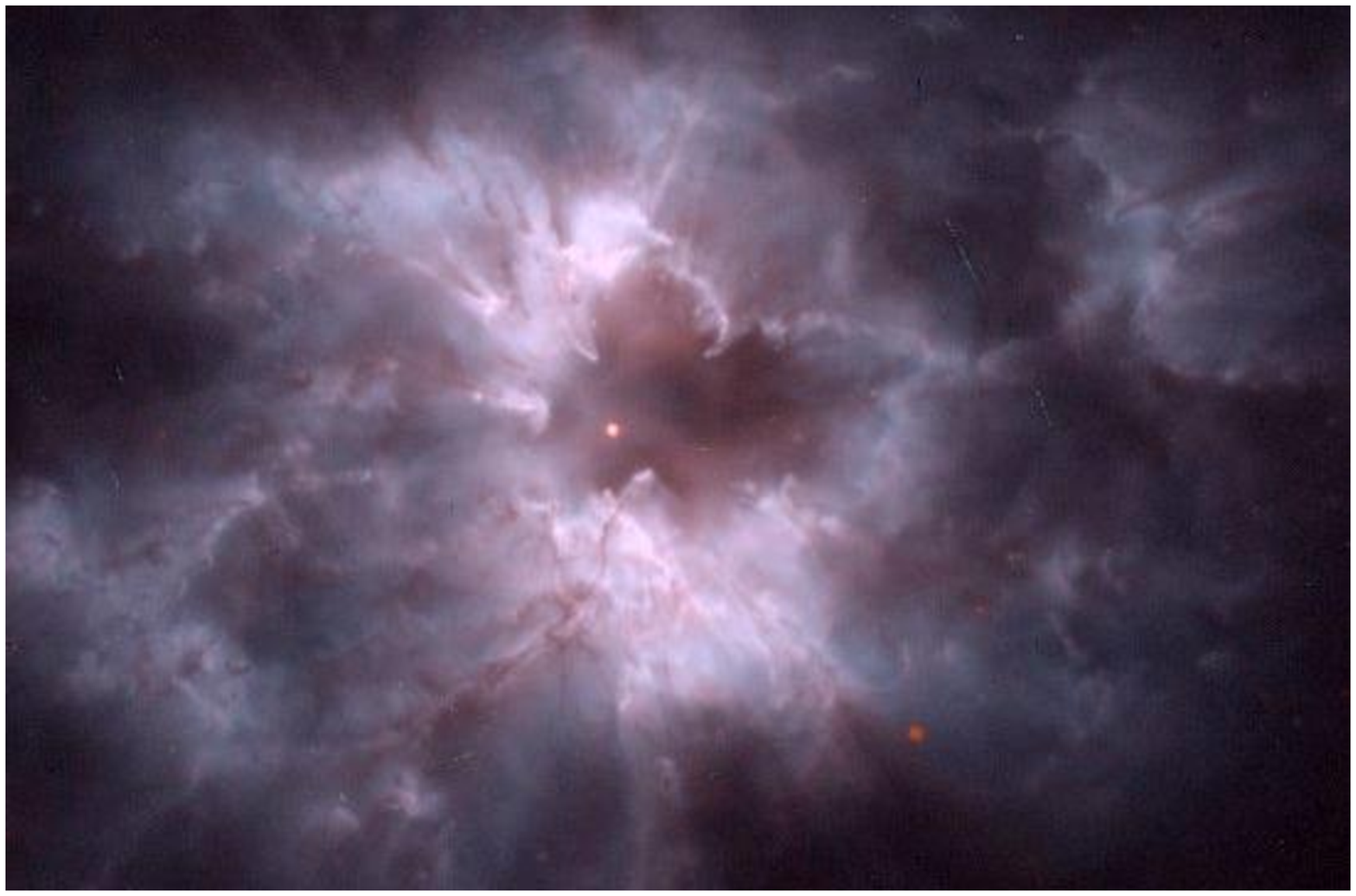}
\vspace{-0.5cm} 
\end{figure}

The equations governing the structure of an active star are given by -
\textcolor{blue}{
\ber
  \frac{dP(r)}{dr} &=& - \frac{G M(r) \rho(r)}{r^2},\, \, \, \mbox{hydrostatic equilibrium}; \\
\frac{dM(r)}{dr} &=& 4 \pi r^2 \rho(r),\, \, \, \mbox{mass-radius relation}; \\
P             &=& P(\rho, T),\, \, \, \mbox{equation of state}; 
\eer
}
\vspace{-1.0cm}
\ber
\frac{dL(r)}{dr} &=& 4 \pi r^2 \rho [\epsilon_n(r) - \epsilon_\nu(r)],\, \, \, \mbox{energy generation}; \\
\frac{dT(r)}{dr} &\propto& - \frac{\rho(r)}{T(r)^3} \frac{L(r)}{4 \pi r^2},\, \, \, \mbox{energy transport} \,;
\eer
where $T$ is the temperature and $L$, $\epsilon_n$, $\epsilon_\nu$ are
the the total, the nuclear and the neutrino luminosities respectively.
The active  phase ends  when a  star runs  out of  its nuclear  fuel -
either because the star never reaches the temperature required to fuse
the next element (low mass stars) or  it reaches the end of the fusion
chain by producing  Fe$^{56}$ (for stars with $M  > 8-10$~\msun). When
low  mass stars  stop burning  their  nuclear fuel  they typically  go
through a red-giant  and a planetary nebula phase and  ends up with no
further energy  generation. As the  residual heat is radiated  away it
cools down and  reaches a temperature low  enough for it to  go into a
quantum  degenerate  phase  (Fig.~\ref{f_pnebula}). For  this  stellar
remnant, the structure  is governed only by the  first three equations
above and  the most important ingredient  is the equation of  state of
the degenerate matter.

\begin{figure}
  \caption{Mass-Radius relation of purely electron-degenerate white dwarfs.
    The dotted curve is for calculations that are inclusive of general relativistic
    corrections. Picture courtesy - {\em Prasanta Bera (IUCAA, Pune)}.}
  \label{f_wdmr}
\vspace{-0.5cm} 
\centering\includegraphics[width=9.0cm]{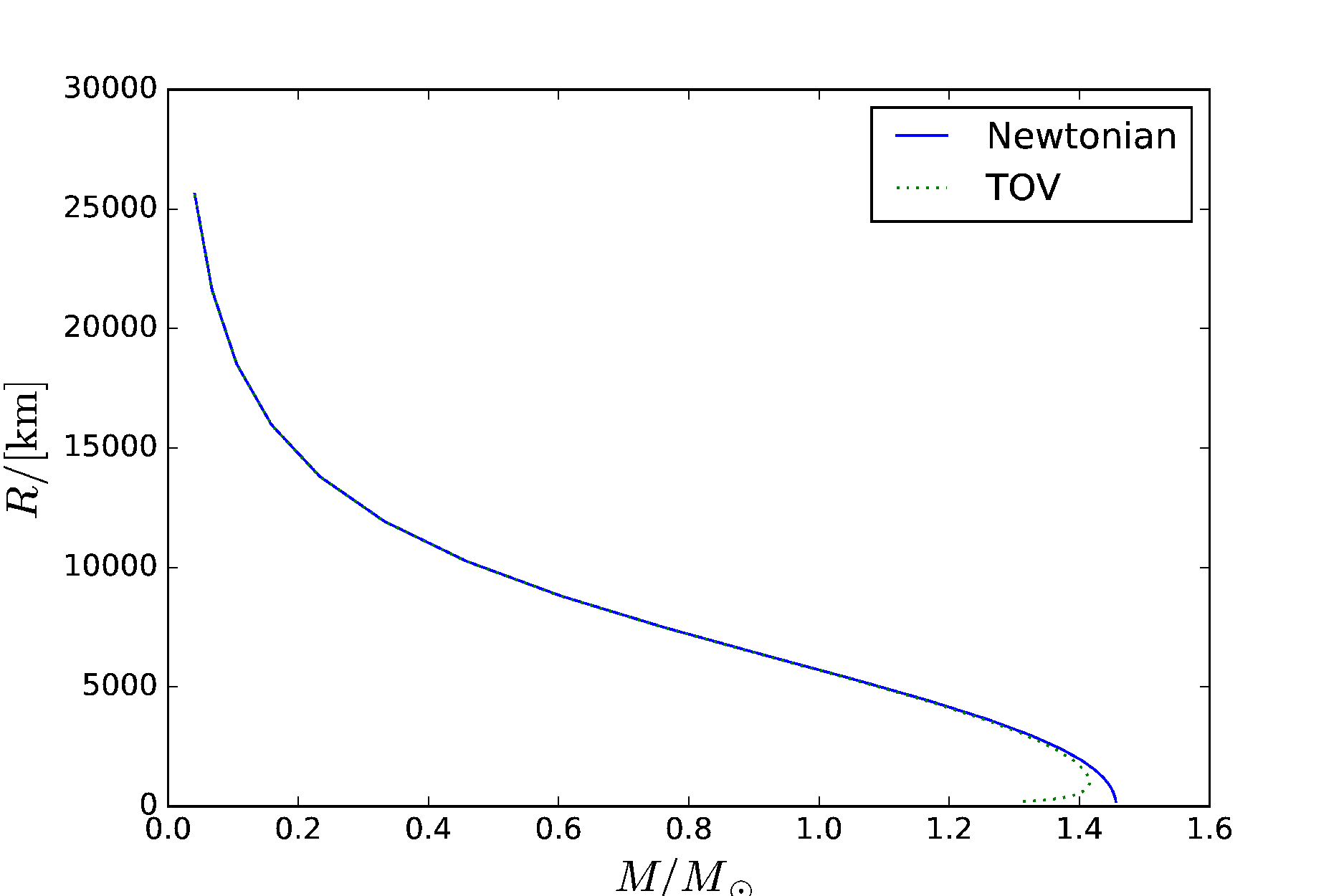}
\vspace{-0.5cm} 
\end{figure}

Table~\ref{t_rmnt} shows stars with different initial masses ending up
as white dwarfs  with different compositions. For all  of them though,
the pressure comes  from the degenerate elections.   However, it needs
to be remembered that with increasing  density the Fermi energy of the
electrons  increase. When  the  density reaches  $\sim 10^6$~\gcc  the
electron  momenta  are   high  enough  for  them  to   be  treated  as
relativistic. In  that case, the  relation between the energy  and the
momentum changes from  $E = p^2/2m$ for  non-relativistic particles to
$E  =   \sqrt{p^2  c^2  +   m^2  c^4}$  for  the   relativistic  case.
Consequently, the pressure is now given by -
\beq
P = \frac{8 \pi m^4 c^5}{3 h^3} \Phi(x) \,,
\eeq
where $x = p_{\rm F}/mc^2$, $p_{\rm F}$ being the Fermi momentum.  The
function $\Phi(x)$ is given by -
\beq
\Phi(x) = \frac{1}{8 \pi^2} \left( x \sqrt{1+x^2} \, (2 x^2/3 - 1) + \ln \left[x + \sqrt{1 + x^2}\right] \right) \,. 
\eeq
Using this equation of state, the structure and the mass-radius relation of
white dwarfs is calculated, as has been shown in Fig.~\ref{f_wdmr}.

The most interesting consequence of the above equation of state is the
case  of  ultra-relativistic  electrons.   In the  limit,  $p_{\rm  F}
\rightarrow  mc$  we  have   $P  \propto  \rho^{4/3}$.  Following  the
arguments, of section~1.2.2 of the previous article in this series, we
arrive at the startling result of
\beq
M = \mbox{\underline{constant}}\,,
\eeq
which is indicated  by the mass-radius curve meeting  the mass-axis at
zero radius.  The  meaning of this is the existence  of an upper limit
for the mass  of a self-gravitating body supported  purely by electron
degeneracy  pressure. This  indeed  is the  famous Chandrasekhar  mass
limit ($M_{\rm  Ch} \simeq 1.44$\msun)  for white dwarfs.  In  1925, a
star as massive  as the Sun but  with radius similar to  the Earth was
discovered. Following R.  H. Fowler's  suggestion that such a star has
densities   in   which  quantum   effects   would   be  important,   S.
Chandrasekhar, then a student  at Presidency College Madras, developed
the theory of white dwarfs.

\subsection{New Directions}

Theoretical research into the interior properties of white dwarfs has
received a boost in recent  times. In particular, two physical effects
have been investigated in detail in the context of its structure. Here
we take a look at these new developments.

\subsubsection{Relativistic Effects}

We have  seen earlier~\cite{konar17b}  that white  dwarfs, made  up of
completely non-relativistic  degenerate electrons, obey  the following
mass-radius relation,
\beq
M R^3 = \mbox{constant} \,,
\label{e_wdmr}
\eeq
implying that  with increasing mass a  white dwarf gets more  and more
compact.  Though, the mass-radius relation  changes from such a simple
one (as seen in Fig.~\ref{f_wdmr}) as the constituent electrons become
relativistic,  the   essential  behaviour   remains  the   same.  With
increasing  compactness  the  effects of  general  relativity  becomes
important  and the  simple  hydrostatic equilibrium  equation used  to
calculate the structure  of a self-gravitating object turns  out to be
inadequate.

The hydrostatic equilibrium equation, modified to incorporate the general
relativistic effects, goes by the name of {\em Tolman-Oppenheimer-Volkoff}
(TOV) equation after the scientists who derived it for the first time and
is given by the following form -
\beq
\frac{dP}{dr}
=- \, \frac{G \left(M(r) + 4\pi r^3 P(r)/c^2 \right)
            \left(\rho (r) + P(r)/c^2\right)}
           {r^2 - 2GM(r)r/c^2}
\eeq
for the structure of a static, spherically symmetric relativistic star
in  isotropic  coordinates.   To   derive  this  equation  from  first
principles is beyond the scope of the present article. However, we can
get  a  sense  of  the  modifications by  realising  that  in  general
relativity, mass and  energy are equivalent quantities and  so are the
mass  density and  energy density  (pressure).  Modification  terms in
TOV, compared  to the simple non-relativistic  hydrostatic equilibrium
equation, arise basically from this concept.

The structure of a relativistic star is typically calculated using the
above TOV  equation. However,  there is  an easy way  to find  out the
significance of the  relativistic effects on a given  star. This can
be done  by quantifying the  compactness in the following  way.  Using
Eq.~\ref{e_wdmr} we find that the  escape velocity from the equatorial
surface of a white dwarf is,
\beq
V_{\rm E} = (2 G M / R )^{1/2} \propto M^{2/3} \,.
\eeq
A comparison of this escape velocity with the velocity of light, gives
us a  measure of the compactness  of the object.  For  a typical white
dwarf of  $M \simeq 1$\msun \,  and $R \simeq 10^4$~KM,  the compactness
parameter turns out to be,
\beq
\frac{V_{\rm E}}{c} \simeq (2 G M/R )^{1/2} \sim 0.02 \,,
\eeq
indicating  that  the  effect   of  relativistic  corrections  to  the
structure of white dwarfs are not  very important. This can be readily
seen   from   Fig.~\ref{f_wdmr}.   The  dotted   curve,   giving   the
mass-relativistic relation  obtained from  the TOV  equation, deviates
only  a  little  from  the mass-relativistic  relation  obtained  from
non-relativistic hydrostatic equation.

\subsubsection{Magnetic Fields}

It is well  known that magnetic field is omnipresent  in the Universe.
In a self-gravitating object the  currents that generate this magnetic
field necessarily flow somewhere inside that object (this is not about
astrophysical objects  in external fields!).  Effectively,  this gives
rise to  a $\jb \times  \bb$ force ($\jb$  - current density,  $\bb$ -
magnetic  field) which  need to  be  balanced by  a redistribution  of
matter.  The structure  calculation of  a self-gravitating  magnetised
object   therefore  requires   a  modification   of  the   hydrostatic
equilibrium equation in the following way -
\beq
\frac{d P_{\rm g}}{dr} + \frac{d P_{\rm m}}{dr} = - \rho(r) g(r),
\label{e_mhydr}
\eeq
where $P_{\rm g}$ and $P_{\rm m}$ are the pressures exerted by the gas
and  the magnetic  field.

\begin{table}[!h]
  \caption{The observed and theoretical maximum value of the magnetic field
    for various astrophysical objects.}
\label{t_mfld}
\vspace{-0.25cm}
\centering
\begin{tabular}{lrrrr} 
& mass & radius & B$^{\rm surface}_{\rm average}$ & B$^{\rm central}_{\rm max}$ \\
& \msun & cm & G & G \\
Earth        & $3 \times 10^{-6}$ & $6 \times 10^8$ & \textcolor{darkblue}{$\lsim 1$}       & \textcolor{darkred}{$2 \times 10^7$}   \\   
Jupiter      & $10^{-3}$          & $7 \times 10^9$ & \textcolor{darkblue}{$\sim 10$}       & \textcolor{darkred}{$5 \times 10^7$}   \\
Brown Dwarf  & $0.01 - 0.1$      & $10^{10}$        & \textcolor{darkblue}{$1 - 10^3$}      & \textcolor{darkred}{$10^7 - 10^8$}     \\
Sun          & 1.0               & $10^{11}$        & \textcolor{darkblue}{$\sim$ few}      & \textcolor{darkred}{$3 \times 10^8$}   \\
White Dwarf  & 1.4 (M$_{\rm Ch}$)  & $10^{9}$         & \textcolor{darkblue}{$10^3 - 10^9$}   & \textcolor{darkred}{$4 \times 10^{12}$} \\ 
Neutron Star & $\lsim 2.0$       & $10^{6}$         & \textcolor{darkblue}{$10^8 - 10^{15}$} & \textcolor{darkred}{$5 \times 10^{18}$} \\
\end{tabular}
\end{table}

The above formulation automatically constrains the maximum strength of
the magnetic field for which a self-gravitating configuration would be
stable.   Table~\ref{t_mfld} shows  the observed  and the  theoretical
maximum  value of  the magnetic  field for  a number  of astrophysical
objects. It is obvious that except  for white dwarfs and neutron stars
the observed value  of the field is  completely insignificant compared
to the  theoretical maximum and hence  plays no role in  modifying the
structure.   On the  other  hand,  much effort  is  being directed  to
understand  the structure  and the  nature of  the magnetic  fields in
white  dwarfs and  neutron stars  at  present. A  series of  excellent
papers by  Prasanta Bera  \& Dipankar  Bhattacharya address  many such
questions (incorporating  general relativistic corrections  and strong
magnetic fields) related to the white dwarfs.

\section*{Acknowledgement}

Technological advancements led  to the discovery of a  large number of
white dwarfs  in recent times, spurring  intense theoretical interest.
As a result, a fortuitous  opportunity to work with Rajaram Nityananda
occurred which allowed  me to learn more about white  dwafs beyond the
mandatory graduate course material.

\end{document}